
\input harvmac
\input epsf

\baselineskip 12pt plus 1pt minus 1pt
\noblackbox

\pageno=0\nopagenumbers\tolerance=10000\hfuzz=5pt
\line{\hfill {\tt hep-ph/9703417}}
\line{\hfill Edinburgh 97/2}
\line{\hfill DFTT 19/97}
\vskip 18pt
\centerline{\bf  ASYMPTOTICALLY FREE PARTONS AT HIGH ENERGY}
\vskip 18pt\centerline{Richard D. Ball\footnote*{\footnotefont
Royal Society University Research Fellow}}
\vskip 14pt
\centerline{\it Department of Physics and Astronomy}
\centerline{\it University of Edinburgh, EH9 3JZ, Scotland}
\vskip 14pt
\centerline{and}
\vskip 14pt
\centerline{Stefano Forte}
\vskip 14pt
\centerline{\it INFN, Sezione di Torino}
\centerline{\it via P. Giuria 1, I-10125 Torino, Italy}
\vskip 36pt
{\narrower\baselineskip 10pt
\centerline{\bf Abstract}
\medskip\noindent
We describe the application of renormalization group improved
perturbative QCD to inelastic lepton-hadron scattering 
at high center-of-mass energy but comparatively low photon virtuality. 
We construct a high energy factorization theorem which
complements the mass factorization theorem used for processes with
high virtualities. From it we derive a renormalization group
equation which resums all large logarithms at high energy, thereby
extending to this regime asymptotic freedom and thus the full
range of perturbative computational techniques. We discuss the 
solution of this equation in various limits, and in particular show 
that the high energy behaviour of physical cross-sections is 
consistent with phenomenological expectations and unitarity bounds.
\smallskip}
\vfill
\line{March 1997\hfill}
\eject \footline={\hss\tenrm\folio\hss}


\def\toinf#1{\mathrel{\mathop{\sim}\limits_{\scriptscriptstyle
{#1\rightarrow\infty }}}}
\def\tozero#1{\mathrel{\mathop{\sim}\limits_{\scriptscriptstyle
{#1\rightarrow0 }}}}
\def\frac#1#2{{{#1}\over {#2}}}
\def\half{\hbox{${1\over 2}$}}\def\third{\hbox{${1\over 3}$}}
\def\quarter{\hbox{${1\over 4}$}}
\def\smallfrac#1#2{\hbox{${{#1}\over {#2}}$}}

\catcode`@=11 
\def\slash#1{\mathord{\mathpalette\c@ncel#1}}
 \def\c@ncel#1#2{\ooalign{$\hfil#1\mkern1mu/\hfil$\crcr$#1#2$}}
\def\lsim{\mathrel{\mathpalette\@versim<}}
\def\gsim{\mathrel{\mathpalette\@versim>}}
 \def\@versim#1#2{\lower0.2ex\vbox{\baselineskip\z@skip\lineskip\z@skip
       \lineskiplimit\z@\ialign{$\m@th#1\hfil##$\crcr#2\crcr\sim\crcr}}}
\catcode`@=12 

\def\PR{{\it Phys.~Rev.~}}
\def\PRL{{\it Phys.~Rev.~Lett.~}}
\def\NP{{\it Nucl.~Phys.~}}
\def\NPBPS{{\it Nucl.~Phys.~B (Proc.~Suppl.)~}}
\def\PL{{\it Phys.~Lett.~}}

\def\NC{{\it Nuov.~Cim.~}}

\def\SPJETP{{\it Sov.~Phys.~J.E.T.P.~}}

\def\vol#1{{\bf #1}}\def\vyp#1#2#3{\vol{#1} (#2) #3}

\def\rg{renormalization group}
\def\trge{transverse renormalization group equation}
\def\lrge{longitudinal renormalization group equation}
\def\pdf{parton distribution function}
\def\as{\alpha_s}
\def\bas{\bar\alpha_s}
\def\asmu{\alpha_s(\mu^2)}
\def\asQ{\alpha_s(Q^2)}
\def\asS{\alpha_s(S^2)}
\def\Lam{\Lambda}
\def\muderiv{\mu^2\frac{\partial}{\partial\mu^2}}

\def\Sderiv{S^2\frac{\partial}{\partial S^2}}


\nref\af{F.~Wilczek {\tt hep-th/9609099}, and ref. therein.}
\nref\CWFGL{H.~Cheng and T.T.~Wu, \PRL\vyp{24}{1970}{759, 1456};
\PR\vyp{D1}{1970}{2775}\semi
G.V.~Frolov, V.N.~Gribov and L.N.~Lipatov, \PL\vyp{B31}{1970}{34}.}
\nref\CW{H.~Cheng and T.T.~Wu, ``Expanding Protons: Scattering at High
Energies'' (MIT Press, 1987) and ref. therein.}
\nref\VV{G.~'t~Hooft, \PL\vyp{B198}{1987}{61}\semi
R.~Jackiw, D.~Kabat and M.~Ortiz, \PL\vyp{B277}{1992}{148}\semi
H.~Verlinde and E.~Verlinde, {\tt hep-th/9302104}.}
\nref\FKL{V.S.~Fadin, E.A.~Kuraev and L.N.~Lipatov,
\PL\vyp{B60}{1975}{50};\hfill\break \SPJETP\vyp{44}{1976}{443}; \vyp{45}{1977}{199}.}
\nref\Lip{L.N.~Lipatov, in ``Perturbative QCD'', ed. A.H.~Mueller (World
Scientific, 1989) and ref. therein.}
\nref\Romsum{R.D.~Ball and A.~DeRoeck, {\tt hep-ph/9609309}, and ref. therein.}
\nref\Froi{M.~Froissart, \PR\vyp{123}{1961}{1053}\semi
A.~Martin, \NC\vyp{42}{1966}{930}; \vyp{44}{1966}{1219}.}
\nref\APe{G.~Altarelli and G.~Parisi, \NP\vyp{B126}{1977}{298}.}
\nref\EGMPR{R.K.~Ellis et al, \PL\vyp{78B}{1978}{281};
\NP\vyp{B152}{1979}{285}.}
\nref\CH{S.~Catani, M.~Ciafaloni and F.~Hautmann, \PL\vyp{B307}{1993}{147}\semi
S.~Catani and F.~Hautmann, \PL\vyp{B315}{1993}{157}; \NP\vyp{B427}{1994}{475}.}
\nref\CFP{G.~Curci, W.~Furmanski and R.~Petronzio, \NP\vyp{B175}{1980}{27}.}
\nref\Weinberg{S.~Weinberg, \PR\vyp{118}{1960}{838}, and in ``The
Quantum Theory of Fields'' (Cambridge, 1996).}
\nref\xLip{S.~Catani et al, \NP\vyp{B336}{1990}{18}; \vyp{B361}{1991}{645}\semi
A.H.~Mueller, \NP\vyp{B415}{1994}{373}\semi
V.~Del~Duca, {\it Scientifica Acta} \vyp{10}{1995}{91}
({\tt hep-ph/9503226})\semi
I. Balitskii, \NP\vyp{B463}{1996}{99}\semi
J.~Jalilian-Marian et al., {\tt hep-ph/9701284} .}
\nref\CQz{M.~Ciafaloni, \PL\vyp{B356}{1995}{74}.}
\nref\CCH{S.~Catani, M.~Ciafaloni and F.~Hautmann,
\PL\vyp{B242}{1990}{97};\hfill\break \NP\vyp{B366}{1991}{135}.}
\nref\CRom{S.~Catani, {\tt hep-ph/9608310}.}
\nref\DGPTWZ{A.~De~R\'ujula et al., \PR\vyp{10}{1974}{1649}.}
\nref\DAS{R.D.~Ball and S.~Forte,
\PL\vyp{B335}{1994}{77}; \vyp{B336}{1994}{77}.}
\nref\Jaro{T.~Jaroszewicz, \PL\vyp{116B}{1982}{291}.}
\nref\Summing{R.D.~Ball and S.~Forte, \PL\vyp{B351}{1995}{313}; 
\vyp{B359}{1995}{362}.}
\nref\LSC{R.K.~Ellis et al, \PL\vyp{B348}{1995}{582}\semi
R.D.~Ball and S.~Forte, \PL\vyp{B358}{1995}{365}\semi
J.~Bl\"umlein et al, \NPBPS\vyp{51C}{1996}{30}\semi
R.S.~Thorne, \PL\vyp{B392}{1997}{463}; {\tt hep-ph/9701241}.}
\nref\scheme{S.~Forte and R.D.~Ball, {\tt hep-ph/9507211}.}
\nref\DSV{R.D.~Ball and S.~Forte, {\tt hep-ph/9607291}\semi
I.~Bojak and M.~Ernst, {\tt hep-ph/9702282}.}
\nref\Mont{S.~Forte and R.D.~Ball, {\tt hep-ph/9610268}\semi
L.~Mankiewicz, A.~Saalfeld and T.~Weigl, {\tt hep-ph/9612297}.}


Asymptotic freedom~\af\ in deep inelastic processes, derived using mass
factorization and the renormalization group, lies at the heart of our
understanding of perturbative QCD. Attempts to use perturbative QCD
to understand processes with high centre-of-mass energies but
relatively low virtualities are as old~\refs{\CWFGL,\CW}, 
but have never had as firm a theoretical foundation. In these
processes distances from the light-cone are no longer short,
but longitudinal distances are compressed by the large relative
longitudinal momenta of the interacting particles~\VV.
At leading order in perturbation theory cross-sections are
asymptotically flat, while at higher orders logarithms of energy (or
rapidity) generate a rise. However these large logarithms eventually
spoil the perturbative expansion. Attempts to sum them~\refs{\FKL,\Lip} 
lead to cross-sections which seem to rise too quickly to
agree with current data~\Romsum, and eventually
violate unitarity bounds~\Froi.

Here we describe an approach to these processes based on a high energy
factorization theorem for inelastic lepton-hadron scattering which, when 
combined with the renormalization
group, implies that QCD becomes asymptotically free at high energies.
The large energy logarithms can then be resummed to give a
self-consistent perturbative description of inelastic scattering at
high energy. A partonic picture is developed similar in many ways to
the conventional one \APe\ useful at high virtualities. We find that
cross-sections then rise only logarithmically, consistent both with
phenomenological expectations and the behaviour of unitarity bounds. 
In this paper we will outline our main results; detailed proofs and 
a wider range of applications will be presented in a forthcoming publication.

Perturbative logarithmic corrections to cross-sections 
arise (in physical gauges) from two particle 
reducible diagrams: two particle irreducible diagrams can be at
most constant at high scales in a nonabelian gauge theory because they
contain no logarithmic 
singularities~\EGMPR. The two particle reducible contributions to
an inelastic virtual photon-hadron cross-section satisfy 
the fundamental factorization~\CH:
\eqn\ktfac{
\bar\sigma^{(2)}(x,Q^2/\Lam^2)=\int_x^1 \frac{dy}{y}
\int_0^{\quarter W^2}\frac{dk^2}{k^2}
C\Big(\frac{y}{x},\frac{Q^2}{k^2};\asmu\Big)
\bar f\Big(y,k^2;\mu^2\Big),}
where, if $q$ and $p$ are the momenta of the virtual photon and hadron,
$Q^2=-q^2$, $x=Q^2/2p\cdot q$, $\Lam$ is a fixed scale,
$\mu$ is the renormalization scale, and parton indices have been suppressed.
This factorization may be proven formally in axial gauges by a projection
argument~\CH\ which generalises that used
to prove mass factorization~\refs{\EGMPR,\CFP}.
The content of eq.~\ktfac\ is that all collinear
singularities can be factored into the 
unintegrated two particle irreducible \pdf\ $\bar f$ in
such a way that the process-independent two particle reducible coefficient
function $C$ only depends on the renormalization scale through the strong
coupling. The dependence on the parton and hadron momenta is
then fixed by simple kinematics.     
Here we will  take~\ktfac\ as our starting point, show how to
derive mass factorization from it and then, by a parallel
argument, derive the energy factorization theorem which will be
the basis for our high energy perturbative expansion.

To expose the symmetry implicit in~\ktfac, we introduce
$S^2=\Lam^2/x$, $l^2=\Lam^2/y$, so that
\eqn\doublefac{
\sigma^{(2)}\Big(\frac{S^2}{\mu^2},\frac{Q^2}{\mu^2};\mu^2\Big)=
\int_0^\infty \frac{dl^2}{l^2}\int_0^\infty \frac{dk^2}{k^2}
C\Big(\frac{S^2}{l^2},\frac{Q^2}{k^2};\asmu\Big)
f\Big(\frac{l^2}{\mu^2},\frac{k^2}{\mu^2};\mu^2\Big),}
where the full dependence on the renormalization scale $\mu^2$
is displayed explicitly (though that of the cross-section is of course
fictitious). The formal extension of the integration limits is of no
consequence for the applications we shall consider (provided the
definitions of coefficient functions and parton distributions are
suitably extended).
Just as the high virtuality limit corresponds to
$Q^2\to\infty$ at fixed $S^2$, so the high energy limit is now
$S^2\to\infty$ at fixed $Q^2$.
Both the convolutions may be undone by taking Mellin transforms:
defining for example
\eqn\melldef{
\sigma_{NM}^{(2)}(\mu^2) \equiv
\int_0^\infty \frac{dS^2}{S^2}\Big(\frac{\Lam^2}{S^2}\Big)^N
\int_0^\infty \frac{dQ^2}{Q^2}\Big(\frac{\Lam^2}{Q^2}\Big)^M
\sigma\Big(\frac{S^2}{\mu^2},\frac{Q^2}{\mu^2};\mu^2\Big),}
the double factorization~\doublefac\ assumes the simple form
\eqn\melldoubfac{
\sigma_{NM}^{(2)}=C_{NM}(\asmu)f_{NM}(\mu^2).}

We next explain how mass factorization can be derived from the double
factorization~\doublefac. At large $Q^2$ we may write~\doublefac\ as
\eqn\tvtwistproj{
\sigma_N^{(2)}(Q^2/\mu^2;\mu^2)
=\frac{1}{2\pi i}\int_{\epsilon-i\infty}^{\epsilon+i\infty} \!\! dM
\Big(\frac{Q^2}{\Lam^2}\Big)^M C_{NM}(\asmu)f_{NM}(\mu^2),}
with the integration contour passing to the right of the origin
and closed on the left. The singularities in the $M$-plane
at any fixed order in perturbation theory may be found by
power-counting arguments~\Weinberg, since the Mellin
transform of $(\Lambda^2/Q^2)^k\ln^n{\Lambda^2\over Q^2}$
is ${n!/(M-k)^{n+1}}$. It follows that the leading singularity at large
$Q^2$ is at $M=0$, singularities further to the left giving
rise to higher twist terms which vanish as powers of
$1/Q^2$. Similarly all two 
particle irreducible contributions to the cross-section are down
by powers of $1/Q^2$, and can thus be ignored at leading twist.

Hence at large $Q^2$ and leading twist we are interested only in the
behaviour of $C_{NM}$ and $f_{NM}$ in the neighbourhood of $M=0$. 
The two particle irreducible 
unintegrated \pdf\ is regular at $M=0$ (again by power
counting~\Weinberg), while the (two particle reducible) 
coefficient function contains
logarithmic singularities which correspond to multiple poles at $M=0$.
We may thus perform a Laurent expansion of the coefficient function
\eqn\tvcoeffexp{
C_{NM}(\asmu) = \sum_{m=-\infty}^\infty C_N^m(\asmu) M^{-m-1},}
so that at large $Q^2$
\eqn\tvcoefflogs{
C_N(Q^2/k^2;\mu^2)=
\sum_{m=0}^\infty \frac{1}{m!} C_N^m(\asmu) \big(\ln Q^2/k^2\big)^m
+O(k^2/Q^2).}
Since the logarithmic singularities are generated perturbatively, the
coefficient $C_N^m(\as)$ will begin at $O(\as^m)$ when $m>0$. Similarly 
the unintegrated \pdf\ has a Taylor expansion
\eqn\tvpdfexp{
f_{NM}(\mu^2) = \sum_{m=0}^\infty f_N^m(\mu^2) M^m,}
where
\eqn\tvpdfmom{
f_N^m(\mu^2) = \frac{1}{m!}\int_0^\infty \frac{dk^2}{k^2}
\big(\ln \mu^2/k^2\big)^m f_N(k^2/\mu^2;\mu^2).}
Now, contributions to the contour integral in eq.~\tvtwistproj\ with a
given power of $\ln {Q^2\over \Lambda^2}$ 
will pick up the coefficients of multiple 
poles in the product of the expansions~\tvcoeffexp\ and~\tvpdfexp. 
It follows that going beyond leading order in~\tvpdfexp\ introduces 
extra powers of $M$, which must be compensated by extra powers of 
$1/M$ in~\tvcoeffexp, and thus extra powers of $\as$. Hence in a 
leading logarithm
approximation we need retain only the first term in \tvpdfmom: writing
the integrated \pdf\ $F_N(\mu^2)=f_N^0(\mu^2)$ the double factorization
\tvtwistproj\ of two particle reducible contributions then becomes simply
\eqn\massfact{
\sigma_N(Q^2/\mu^2;\mu^2) = C_N(Q^2/\mu^2;\asmu) F_N(\mu^2) + O(1/Q^2),}
i.e. the mass factorization of the complete cross-section at high
$Q^2$.

Renormalization group invariance (i.e. $\mu^2$ independence) 
of the cross-section then implies that if
\eqn\tvevfnz{
\muderiv F_N(\mu^2) = \gamma_N(\asmu)F_N(\mu^2),}
the cross-section may be written as
\eqn\massevol{
\sigma_N(Q^2/\mu^2;\mu^2) = C_N(1;\asQ) \Gamma_N(Q^2,\mu^2) F_N(\mu^2) +
O(1/Q^2),}
where $C_N(1;\as)=C_N^0(\as)$, and the evolution factor
\eqn\tvevfac{
\Gamma_N(Q^2,\mu^2)\equiv \exp \int_{\asmu}^{\asQ}
\frac{d\alpha}{\beta(\alpha)}\gamma_N(\alpha),}
with $\beta(\asmu)\equiv \partial\asmu/\partial\ln\mu^2$. All
the dependence on $\mu^2$  may then be eliminated explicitly by 
setting $F_N(Q^2)=\Gamma_N(Q^2,\mu^2) F_N(\mu^2)$. 

This derivation of leading-order mass factorization may be
extended to all orders in perturbation theory by choosing suitable mass
factorization schemes (where, in particular, all terms in the
expansion eq.~\tvpdfexp\ have the same $\mu^2$ dependence). 
Eqs.~\massfact-\tvevfac\ then still hold but
with
\eqn\tvintpdfdef{
F_N(Q^2)\equiv f_N^0(Q^2)+C_N^{0}(\asQ)\inv\sum_{m=1}^\infty
C_N^{m}(\asQ)f_N^m(Q^2).}
Beyond leading order this expression involves increasingly
higher logarithmic moments~\tvpdfmom\ of the unintegrated \pdf, so
that it is in principle not possible to reconstruct the unintegrated \pdf\ from
the integrated.

Our main interest in this derivation is that because of the symmetry
between $Q^2$ and $S^2$ we may use a very similar argument to derive
an energy factorization theorem valid at high energies, i.e. as
$S^2/\Lam^2\to\infty$, rather than at high virtualities. In the high
energy limit two particle irreducible graphs 
are no longer power suppressed. However they are still
free of logarithmic singularities~\EGMPR, at least at a fixed order in
perturbation theory, and thus in the ultraviolet they are at most constant.
Graphs which grow logarithmically are thus two particle 
reducible (in either the $s$ or
$t$ channel), and in fact, as can be shown by explicit calculation,
must be two gluon reducible. This means that we can put all the
two quark reducible and 
two particle irreducible graphs together into a contribution to the
cross-section which, though not perturbatively
calculable, behaves as a constant in
the high energy limit (see fig.~1).
The remaining two gluon reducible contributions are then given by
\eqn\lgtwistproj{
\sigma_M^{(2)}(S^2/\mu^2;\mu^2)
=\frac{1}{2\pi i}\int_{\epsilon-i\infty}^{\epsilon+i\infty} \!\! dN
\Big(\frac{S^2}{\Lam^2}\Big)^N C_{NM}(\asmu)f_{NM}(\mu^2),}
where again the contour passes just to the right of the origin and is
closed on the left, and the rightmost singularity is that at $N=0$,
the remaining singularities giving contributions which are power
suppressed as $S^2/\Lam^2\to\infty$.

It follows that the leading behaviour at high energies is found
expanding $C_{NM}$ and $f_{NM}$ in
the neighbourhood of $N=0$. All the logarithmic singularities are
again in the coefficient function: 
the \pdf\ is two particle irreducible and thus has at most
a simple pole at $N=0$, so we expand it as
\eqn\lgpdfexp{
f_{NM}(\mu^2) = N\inv \sum_{n=0}^\infty f_M^n(\mu^2) N^n,}
so that
\eqn\lgpdfmom{\eqalign{
f_M^n(\mu^2) = \frac{1}{n!}\int_0^\infty \frac{dl^2}{l^2}
\big(\ln \mu^2/l^2\big)^n l^2\frac{\partial}{\partial\l^2}f_M(l^2/\mu^2;\mu^2)
= -\frac{1}{n!}\int_0^1 dy
\big(\ln y\big)^n \frac{\partial}{\partial y}\bar f_M(y;\mu^2).}}
and in particular for $n=0$ we have simply
\eqn\lgpdfzmom{
f_M^0(\mu^2)=\lim_{y\to 0} \bar f_M(y;\mu^2).}
The coefficient function admits an expansion analogous to~\tvcoeffexp:
\eqn\lgcoeffexp{
C_{NM}(\asmu) = N\sum_{-\infty}^\infty C_M^n(\asmu) N^{-n-1},}
so that
\eqn\lgcoefflogs{
C_M(S^2/l^2;\mu^2)\toinf{S^2/l^2}
\sum_{n=0}^\infty \frac{1}{n!} C_M^{n+1}(\asmu) \big(\ln S^2/l^2\big)^n
+O(l^2/S^2),}
where $C_M^n(\as)$ begins at $O(\as^{n})$ when $n>0$.
Because the top rung is necessarily a quark, which cannot generate a 
logarithm of $S^2$, there is one more power of $\as$ in this expansion 
than in the similar expansion~\tvcoeffexp.

Energy factorization now follows in precisely the same way as mass
factorization: in the leading logarithm approximation we need retain
only the first term in \lgpdfexp: calling this $F_M(\mu^2)$ we
immediately get
\eqn\energyfact{
\sigma_M(S^2/\mu^2;\mu^2) = C_M(S^2/\mu^2;\asmu) F_M(\mu^2)+
\sigma^q_M(S^2/\mu^2;\mu^2)+O(1/S^2),}
where $\sigma^q_M$ contains the two gluon irreducible terms [terms (b)
and (c) in fig.~1] and approaches a constant in the high energy
limit, while  the first term [term(a) in fig.~1] grows logarithmically with $S^2$. 

We may now again use \rg\
invariance of the cross-section to derive the longitudinal evolution equation
\eqn\lgevintpdf{
\muderiv F_M(\mu^2) = \gamma_M(\asmu)F_M(\mu^2),}
where $\gamma_M(\asmu)$ is a new longitudinal anomalous dimension
(as opposed to the conventional transverse anomalous dimension
$\gamma_N(\asmu)$ defined in \tvevfnz). The energy
factorization \energyfact\ then becomes
\eqn\energyevol{
\sigma_M(S^2/\mu^2;\mu^2) = C_M(1;\asS) \Gamma_M(S^2,\mu^2) F_N(\mu^2)
+ \sigma^q_M(S^2/\mu^2;\mu^2)+ O(1/S^2),}
where $C_M(1;\as)=C_M^0(\as)$, and the evolution factor
\eqn\tvevfac{
\Gamma_M(S^2,\mu^2)\equiv \exp \int_{\asmu}^{\asS}
\frac{d\alpha}{\beta(\alpha)}\gamma_M(\alpha).}
Again these results may be extended to all orders in perturbation
theory by choosing a suitable class of factorization schemes: the longitudinal
integrated \pdf\ is then given by
\eqn\lgintpdfdef{
F_M(S^2)\equiv f_M^0(S^2)+C_M^{0}(\asS)\inv\sum_{n=1}^\infty
C_M^{n}(\asS)f_M^n(S^2),}
so that again the unintegrated \pdf\ cannot be
reconstructed from it. It follows that the two integrated \pdf s
$F_N(\mu^2)$ and $F_M(\mu^2)$ are not directly related to one another.

Just as the usual evolution eq.~\massevol\ resums logarithms of the 
transverse scale $Q^2$, the evolution equation~\energyevol\ 
resums logarithms of the large longitudinal scale $S^2$. 
Thus while the former evolution is asymptotically free at 
high virtualities, the latter is asymptotically free at high 
energies, i.e. at small $x$. In both cases, the resummation 
follows directly from renormalization group
improvement of the dependence on the unphysical
variable $\mu^2$: the scale of this dependence is then necessarily the large
physical scale which underpins the appropriate
factorization formula, eq.~\massfact\ and \energyfact\ respectively.
An inappropriate choice of renormalization scale in either case, 
such as for example $Q^2$ in a high energy process, results in a
badly behaved perturbative approximation, signalled by a large
scheme dependence of the results, which is due to the fact that large 
logarithms are not being resummed. 
Both evolutions require nonperturbative input: in
the former case a starting distribution
$F_N(Q_0^2)$, and in the latter  case a starting distribution
$F_M(S_0^2)$. However, the renormalization group improved
factorizations \massevol\ and \energyevol\ 
imply that the final results are independent of the starting
scales, $Q_0^2$ and $S_0^2\equiv\Lam^2/x_0$ respectively.

Before we can consider perturbative evolution in the longitudinal case
we must first know the longitudinal anomalous dimension
$\gamma_M(\as)$, or, rewriting \lgevintpdf\ as
\eqn\lgevintpdfAP{
\Sderiv F(Q^2,S^2)
= \frac{\asS}{2\pi}\int_0^\infty \frac{dk^2}{k^2}P(Q^2/k^2;\asS)F(k^2,S^2),}
the splitting function $P(\kappa;\asS)$ given by the inversion of
the Mellin transform
\eqn\splitdef{\gamma_M(\as)=\frac{\as}{2\pi}\int_0^\infty
\frac{d\kappa}{\kappa}
\kappa^M P(\kappa;\as).}

The splitting function may be calculated in the Weizs\"acker--Williams
approximation by considering the emission of
quasi-collinear partons in the infinite momentum frame (so that the
transverse momenta $k$ are much smaller than the longitudinal momenta
$p$), and picking out the coefficient of whichever large logarithm we are
interested in. In the transverse case \APe\ this is related to
the collinear divergence of a parton propagating in the $t$-channel
with small transverse momenta $k^2$, while in the longitudinal case
it is the infrared divergence due to the emission of a soft parton
of longitudinal momentum $yp$.

When computing the transverse splitting function, it is sufficient to
consider $t$-channel emissions, since emission in the $s$-channel
produces no extra collinear logarithms. However, for the longitudinal
splitting function, both $s$ and $t$ channel emissions can contribute:
for purely gluonic processes the sum of the amplitudes for emission of
$t$ and $s$ channel gluons (see fig.~2) is (suppressing colour factors
for simplicity)
\eqn\stamp{
A^{(1)}_{++}(k,k^\prime)
=2g^2\left({(k^\prime-k)^*\over
|k-k^\prime|^2}{1\over y}{(-k^*)\over z}-{y\over
|k^\prime|^2}\left({-k^*\over x}\right) {{k^\prime}^*\over
y}\right)}
where $k=k_x+ik_y$ and the subscript $++$ denotes the helicities of 
the emitted and final-state gluons: amplitudes for other helicity
combinations are found by complex conjugation
of the pertinent momenta. Evaluating the 
photon-parton cross section in the Weizs\"acker-Williams approximation
in terms of the spin and colour averaged square of the
amplitude~\stamp\ and the total inclusive cross section for the 
photon to scatter off parton $C$
\eqn\wwone
{\eqalign{d\sigma^{(1)}
&={d^3p_B\over (2\pi)^3 2E_B } {d^3p_{B^\prime}
\over (2\pi)^3 2E_{B^\prime} }
|A^{(1)}(k,k^\prime)|^2 {1\over p_C^2} x \sigma(\gamma^* C\to X)\cr
&=d \sigma^{(0)} 
{d^2 {k^\prime}\over \pi}{dy\over y} 2 C_A{\alpha\over 2\pi}
{|k|^2\over |k^\prime|^2}{1\over |k-k^\prime|^2}\cr
&=d \sigma^{(0)} {dy\over y}{d\kappa\over\kappa} 2 C_A {\alpha\over 2\pi}
{1\over|1-\kappa|},\cr}}
where $\kappa= |k'|^2/|k|^2$. 
The photon-parton cross section $d \sigma^{(0)}$ thus acquires 
an extra energy logarithm due to the extra emission.
The real contribution to the splitting function may thus be simply
identified with the coefficient 
$2C_A/|1-\kappa|$ of this logarithm. Virtual gluon emission
produces a contribution to the splitting function  
proportional to $\delta(1-\kappa)$ (because
it leaves the momentum of the propagating parton unchanged),
and regularises the singularity at $\kappa=1$, to give
\eqn\splfc{P(\kappa)=2 C_A\left[ 2{1\over 1-\kappa}\Big|_+\Theta(1-\kappa)-
{\rm P} {1\over 1-\kappa}+c \delta(1-\kappa)\right],}
where $\Theta$ is the step function, P indicates the principal part,
$+$ is the plus prescription defined in ref.~\APe, and $c$ is a finite
real number, which may determined 
to be zero by explicit calculation of the one
loop virtual corrections, or from the knowledge of the two-loop
transverse splitting function, as described below.

The longitudinal anomalous dimension may then be determined from the
splitting function~\splfc\ by taking Mellin moments~\splitdef:
\eqn\longanomdim{
\gamma_M(\as)=C_A\frac{\as}{\pi}\chi(M)+O(\as^2),}
where
\eqn\lipfn{
\chi(M)=\int_0^1 \frac{d\kappa}{1-\kappa}(\kappa^{M-1}+\kappa^{-M}-2)
=2\psi(1)-\psi(M)-\psi(1-M),}
and $\psi(z)$ is the digamma function. 

Solving the evolution equation~\lgevintpdf\ with the anomalous
dimension~\longanomdim\ corresponds to iterating the s- and t-channel 
emissions of fig.~2, with the kinematic constraint of strong ordering of
longitudinal momenta $1/y_1<<\dots<<1/y_n<<1/x$. The iteration may
now be checked by explicit computation: this is one of the advantages
of using the Weizs\"acker-Williams approximation. The diagrams 
generated in this way are then all (cut) gluon ladders 
with rungs dresses by rainbow self-energy corrections 
(plus virtual corrections). These are the same 
leading logarithmic corrections
as those to the unintegrated \pdf\ summed at fixed coupling by the
BFKL equation~\refs{\FKL,\Lip,\xLip}: $\chi(M)$ is the Lipatov
characteristic function and thus the longitudinal splitting 
function $P(\kappa)$  coincides with the kernel of the (forward) BFKL
equation. The evolution equation~\lgevintpdfAP, because it has been
derived from the appropriate factorization theorem \energyfact,
performs instead a renormalization group resummation of these
logarithms which drive the evolution of the integrated \pdf\ $F_M(S^2)$. 
The perturbative approach is then justified self consistently because 
the coupling becomes asymptotically free in the high energy limit.

When quarks are included, it is easy to check by a similar calculation
that only the process of gluon emission from a quark can generate
energy logarithms. Thus at the level of leading logarithms any initial
quark must immediately turn into a gluon, in which case its
contribution may be absorbed into the initial gluon distribution,
or else be struck by the virtual photon, in which case it contributes
to the asymptotically flat term $\sigma_M^q$ in \energyevol.
It follows that at the leading order there is only one type of evolving 
parton, namely the gluon. It is not difficult to see that one can choose
factorization schemes in which this remains true to all orders.

Finally, coefficient functions $C_M(1;\as)$ may be calculated at leading order
by evaluating the quark box for an incoming off-shell gluon, taking
Mellin transforms and setting $N=0$. For the contributions of light
quarks to $F_2$ and $F_L$ this gives
\refs{\CH,\CQz}
\eqn\cofun
{\eqalign{
C^2_M(1;\as)&={\as\over 2\pi} T_R{2(2+3M-3M^2)\over M(3-2M)}
{\Gamma^3(1-M)\Gamma^3(1+M)\over\Gamma(2+2M)\Gamma(2-2M)}+O(\as^2),\cr
C^L_M(1;\as)&={\as\over 2\pi} T_R{4(1-M)\over 3-2M}
{\Gamma^3(1-M)\Gamma^3(1+M)\over\Gamma(2+2M)\Gamma(2-2M)}+O(\as^2);\cr}}
coefficient functions for heavy quark production may be deduced from
the results in ref.~\refs{\CCH,\CRom}. Whereas at leading
order the $M$-dependence of any particular coefficient function 
can be absorbed in a redefinition of the starting gluon 
distribution $G_M(S_0^2)$, and has thus no physical
meaning, ratios of coefficient functions are physically meaningful.

The complementary role of the longitudinal and transverse 
renormalization group equations \lgevintpdf\ and
\tvevfnz\ may be further elucidated noting that they are 
connected by a duality relation. If for simplicity the coupling is 
held fixed and the equations are truncated at leading order, they take the form
\eqn\LOev{
\frac{\partial}{\partial\ln S^2}\sigma_{M}(S^2)
= \bas \chi(M) \sigma_{M}(S^2),\qquad
\frac{\partial}{\partial\ln Q^2}\sigma_{N}(Q^2)
= \bas \phi(N) \sigma_{N}(Q^2),}
where the leading order pure gluon transverse anomalous dimension 
$\gamma^{gg}_N(\as)\equiv \bas\phi(N)$, with
$\bas\equiv\frac{C_A}{\pi}\as$, while the corresponding longitudinal
anomalous dimension is given by \longanomdim. Taking a second Mellin
transform of both equations, we then find that
\eqn\LOevMN{
N \sigma_{NM} = \bas \chi(M) \sigma_{NM},\qquad
M \sigma_{NM} = \bas \phi(N) \sigma_{NM},}
respectively. So both equations can be viewed as
defining a trajectory in the $N$-$M$ plane, each of which approximates
at leading order in $\bas$ the ``true'' trajectory (as given by a 
hypothetical all-order computation), the former when
$N$ is small, the latter when $M$ is small. Both equations will only be
valid together if both $N$ and $M$ are small: since both $\chi$ and
$\phi$ have a simple pole with unit residue at the origin,  both
equations then reduce to $NM=\bas$, which corresponds (at fixed
coupling) to the double logarithmic or double scaling 
approximation~\refs{\DGPTWZ,\DAS}. 

This however also implies that we may turn the
\lrge\ into a \trge, and vice versa, by defining leading singularity
anomalous dimensions implicitly as
\eqn\LSad{
1 =
\smallfrac{\bas}{N}\chi\Big(\gamma^{LS}_T\big(\smallfrac{\bas}{N}\big)\Big),
\qquad
1 =
\smallfrac{\bas}{M}\phi\Big(\gamma^{LS}_L\big(\smallfrac{\bas}{M}\big)\Big),}
in terms of which the two evolution equations \LOevMN\ may  be
written as
\eqn\LSev{
M \sigma_{NM} = \gamma^{LS}_T\big(\smallfrac{\bas}{N}\big) \sigma_{NM},\qquad
N \sigma_{NM} = \gamma^{LS}_L\big(\smallfrac{\bas}{M}\big) \sigma_{NM},}
respectively. Thus the leading singularities in $N$ can also be summed
up explicitly using a \trge, provided the anomalous dimension 
$\gamma_N(\as)=\gamma^{LS}_T(\smallfrac{\bas}{N})$ is used, which
which adds up the leading singularities in $N$ 
in the transverse anomalous dimension to all
orders~\refs{\CH,\Jaro,\Summing}; 
conversely the $M$ singularities may be summed up using
a \lrge.

This duality can thus be used to 
provide powerful cross-checks on the transverse and longitudinal
anomalous dimensions, since knowledge of the leading order
transverse anomalous dimension determines the leading singularities of
the longitudinal anomalous dimension to all orders, and conversely. 
For example the result $c=0$ in \splfc\ is
equivalent to the vanishing of the leading singularity at NLO
(i.e. the $O(\as^2/N^2)$ term in the transverse anomalous dimension~\CFP).
It is now also clear that the failure of attempts~\LSC\ 
to extend the reach of the transverse
evolution equations to high energies~\tvevfnz\ by simply adding
the leading singularities $\gamma^{LS}_T(\smallfrac{\bas}{N})$ to a fixed order
truncation of $\gamma_N(\as)$ is due to the fact that
while the mass singularities are being resummed (because the coupling runs with
$Q^2$), the energy singularities are merely summed~\Summing. This explains the
large scheme dependence in such calculations~\scheme, and ultimately their
failure to agree with experiment~\refs{\Romsum,\DSV}.

We can now discuss some of the consequences of our results for the
behaviour of cross-sections at asymptotically high energies.
The structure function at leading order is determined  by 
solving the
evolution equation~\energyevol\ with the anomalous
dimension~\longanomdim, 
and combining with the appropriate coefficient function~\cofun:
\eqn\efftwo{
F_2(x,Q^2)=\frac{1}{2\pi i}
\int_{\epsilon-i\infty}^{\epsilon+i\infty} \!\! dM
\Big(\frac{Q^2}{\Lam^2}\Big)^M \half\langle e^2\rangle C^2_M(1;\asS)
\Big(\frac{\ln S^2/\Lam^2}{\ln S_0^2/\Lam^2}\Big)^{\gamma^2\chi(M)}
G^0_M +F_2^q(Q^2),}
where $\langle e^2\rangle$ is the average of the light quark charges, 
$\gamma\equiv \sqrt{4C_A/\beta_0}$, and $\beta_0$ is the leading
coefficient of the $\beta$-function.
The a priori unknown inputs at the starting scale
$S_0$ are the initial integrated gluon distribution $G^0(Q^2)$
(the only evolving longitudinal parton distribution)
and the unevolving quark piece
$F_2^q(Q^2)$. To determine the asymptotic behaviour at high energy 
we may take a dipole expression $Q^2/(Q^2+\Lam_0^2)$ for
$G^0(Q^2)$, which has the correct behaviour at both
large and small $Q^2$. Its Mellin transform is then
\eqn\Gzero{
G^0_M = \frac{G_0\pi}{\sin\pi M}\Big(\frac{\Lam^2}{\Lam_0^2}\Big)^M.}

At high energies (and thus large $\eta$) the integral \efftwo\ will be
dominated by the contribution from the saddle point at $M_s$, where
\eqn\saddle{
t+\gamma^2\chi'(M_s)\eta=0,}
where $t\equiv\ln Q^2/\Lam_0^2$ (note that the nonperturbative input sets the
scale for $Q^2$) while
$\eta\equiv\ln\frac{\ln S^2/\Lam^2}{\ln S_0^2/\Lam^2}=\ln\frac{\ln x}{\ln
x_0}$.
Since $\chi'(M)$ increases monotonically as $M$ increases from zero to
one, the precise position of the saddle point will depend on the ratio
$t/\eta$. 

Consider first the high energy limit $\eta\to\infty$, $t/\eta\to 0$ (when
$t$ is fixed this is the so-called Regge limit -- see fig.~3). Then expanding
$\chi'(M)$ about $M=\half$, we have $M_s = \half -
t/28\zeta_3\gamma^2\eta +\ldots$, whence the asymptotic behaviour
\eqn\heasymp{
F_2(x,Q^2)\tozero{x}{\cal N}_0 \left(\frac{Q^2}{\Lam_0^2}\right)^{\half}
\eta^{-\half}
\left(\ln{ 1\over x}\right)^{4\ln 2\gamma^2-1} +F^q_2(Q^2),}
valid in the region $\eta \gg |t|/28\zeta_3\gamma^2$. Subleading
corrections to this
behaviour are down by powers of $|t|/\eta$, while corrections due to
higher orders in perturbation theory are suppressed by powers of $\ln
1/x$.

It follows that in the high energy limit $F_2$ grows logarithmically at a
universal
rate, quite independently of the starting conditions, since eventually
the first term in \heasymp\ must always overcome the
second. Furthermore, in
this limit the $Q^2$ dependence is also universal. 
The universality is a direct consequence of the
fact that their is only one evolving parton at high energies. This
parton can be viewed as playing the role of the pomeron: it generates a
universal rise in the high energy limit. However the rise is
only logarithmic, rather than powerlike, which is 
compatible with the logarithmic nature of unitarity bounds~\Froi. 

Several asymptotic predictions can be derived from this universality, and
its applicability to other structure functions. For example using \cofun\
we see that asymptotically $F_2$ and $F_L$ grow in the same way,
and in fact
\eqn\asymprat{
\lim_{x\to 0}\frac{F_L(x,Q^2)}{F_2(x,Q^2)}= \frac{2}{11}.}
More generally we can derive sum rules
\eqn\sumrules{
\int_0^\infty \frac{dQ^2}{Q^2} (Q^2)^{-M}F_2(x,Q^2)
= \frac{2+3M(1-M)}{2M(1-M)}\int_0^\infty \frac{dQ^2}{Q^2}(Q^2)^{-M}F_L(x,Q^2),}
valid for $0<M<1$. Similar sum rules may be derived for other inclusive
structure functions. Corrections are now down by powers of $\asS$,
rather than powers of $\eta$.

Let us now turn to the opposite limit, i.e. that in which as $\eta\to\infty$,
$|t|/\eta\to\infty$. In the high virtuality limit
$t/\eta\to +\infty$ the position of the
saddle point shifts to just above the origin:
$M_s=\frac{\gamma}{\rho}-2\zeta_3\frac{\gamma^4}{\rho^4}+\ldots$,
where  we write $\sigma\equiv\sqrt{|t|\eta}$ and
$\rho\equiv\sqrt{|t|/\eta}$,
(similarly to~\DAS, but with the roles of $1\over x$ and $Q^2$
interchanged). 
We then find that asymptotically
\eqn\SDAS{
F_2(x,Q^2)\toinf{Q^2}{\cal N}_+
\sigma^{-\half}\rho e^{2\gamma\sigma-\eta}+F^q_2(Q^2),}
valid whenever $\rho\gg(2\zeta_3)^{\third}\gamma$. This is even softer
than the universal high energy behaviour \heasymp. Similarly in the
low virtuality limit $t/\eta\to-\infty$, i.e. the photoproduction
limit (see fig.~3), the saddle point shifts to just below unity:
$M_s=1-\frac{\gamma}{\rho}+2\zeta_3\frac{\gamma^4}{\rho^4}+\ldots$,
and asymptotically
\eqn\photo{
F_2(x,Q^2)\tozero{Q^2}{\cal N}_-
(Q^2/\Lam_0^2)\sigma^{-\half}\rho^2 e^{2\gamma\sigma-\eta}+F^q_2(Q^2).}
The extra factor of $Q^2$ now ensures the finiteness of the
perturbative contribution to the photo-production cross-section,
while the growth with energy is softer than any power of $\ln S^2$,
and thus inside the Froissart bound \Froi, as it should be. For
$F_L$ an extra power of $Q^2$ is generated by the coefficient function
provided that the quark mass is kept finite.

Both the results~\SDAS\ and~\photo\ must however be treated with
caution, since whenever $|t|$ becomes large we start to leave the
the region of validity of the \lrge: logarithms of $Q^2$ become as
important as logarithms of $S^2$. This can be seen in the perturbative
expansion of the anomalous dimension $\gamma_M(\as)$: as $M\to 0$
higher order terms behave as $(\as/M)^m$, generating large ultraviolet
logarithms, while as $M\to 1$ they behave as $(\as/(1-M))^m$,
generating large infrared logarithms, and in both cases leading to the
breakdown of the perturbation series. This is precisely analogous to
the breakdown of the \trge\ as $x\to 0$ and $x\to 1$ respectively.

The contributions from the infrared regions (i.e. the Sudakov region
and the photoproduction region) are in both cases suppressed by the 
fact that they are kinematic boundaries, and thus the \pdf s are small. 
The ultraviolet regions instead are complementary, since
the behaviour \SDAS\ matches smoothly into the double
scaling behaviour \refs{\DGPTWZ,\DAS}. In particular the asymptotic 
behaviours \heasymp\ and \SDAS\ provide a soft boundary
condition to the \trge\ at small $x$ (i.e. rising more slowly
than any power of $1/x$), so as $Q^2$ increases the double scaling  
rise eventually dominates. This explains why double scaling is so successful
phenomenologically \refs{\Romsum,\Mont}.

In conclusion, we have discussed a new factorization theorem, valid at
asymptotically high energies but moderate values of $Q^2$, which
correctly resums all high
energy logarithms. We found that there is then a consistent partonic picture
of high energy scattering in which the partons are asymptotically
free. There is only one type of evolving parton, which may be thought
of as a gluon. As it evolves it generates a universal growth of structure
functions at small $x$ reminiscent of that traditionally described by
the pomeron, but much softer (logarithmic), and thus consistent with
known unitarity limits. This might have far reaching implications
not only for our understanding of other inelastic lepton-hadron processes (such as
diffractive and rapidity gap events), but also for photon-photon  
and purely hadronic processes.

\bigskip
{\bf Acknowledgements}: We would like to thank S.~Catani, V.~Del~Duca
and E.~Predazzi for various discussions during the course of this
work. We also thank the theory division at CERN where part of this
work was done, and the PPARC for a Visiting Fellowship award.

\vfill\eject
\listrefs
\vfill\eject
\topinsert
\vbox{\hbox{
\hfil\epsfysize=5.5truecm\epsfbox{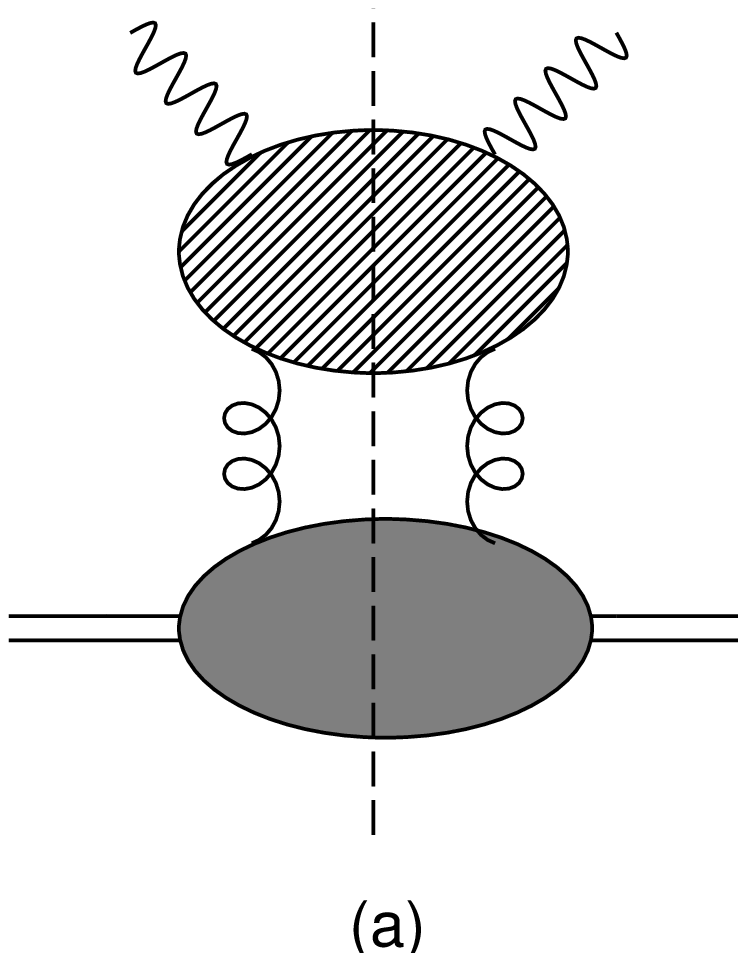}\hskip 1truecm
\vbox{\vskip -.25truecm
\epsfysize=4.8truecm\epsfbox{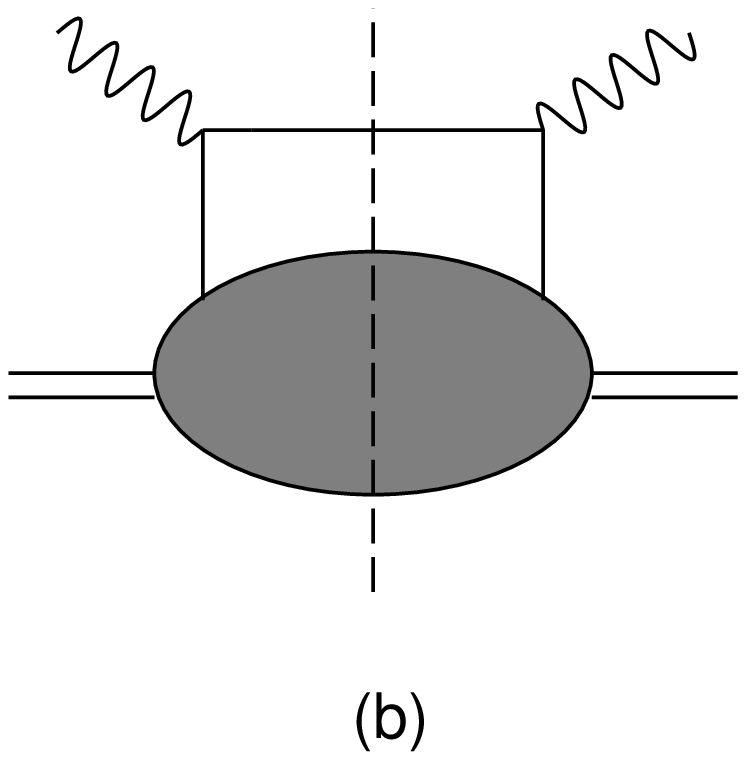}}\hfil\hskip-.4truecm
\vbox{\vskip -.1truecm
\epsfysize=4.5truecm\epsfbox{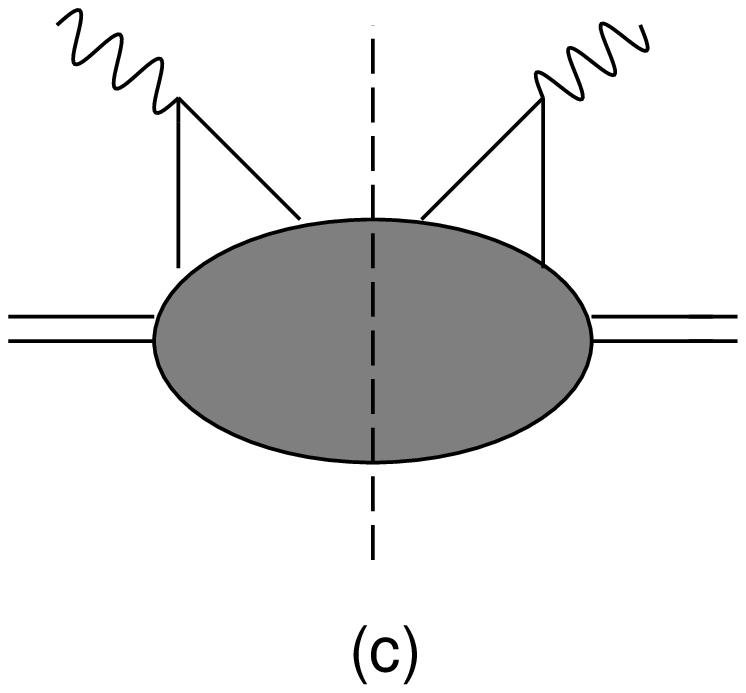}}\hfil}
\bigskip\noindent{\footnotefont\baselineskip6pt\narrower
Figure 1: Contributions to the forward amplitude: a) two gluon
reducible, b) two gluon irreducible but two quark reducible
c) two particle irreducible. At high $Q^2$ a) and b) are leading
twist and evolve, while c) is higher twist (power suppressed). At high
$S^2$ none of them are power suppressed, but only a) evolves, while b)
and c) are asymptotically constant.
\medskip}}
\endinsert
\topinsert
\vbox{\vskip3truecm \hbox{\hskip 2truecm
\hfil\epsfysize=5truecm\epsfbox{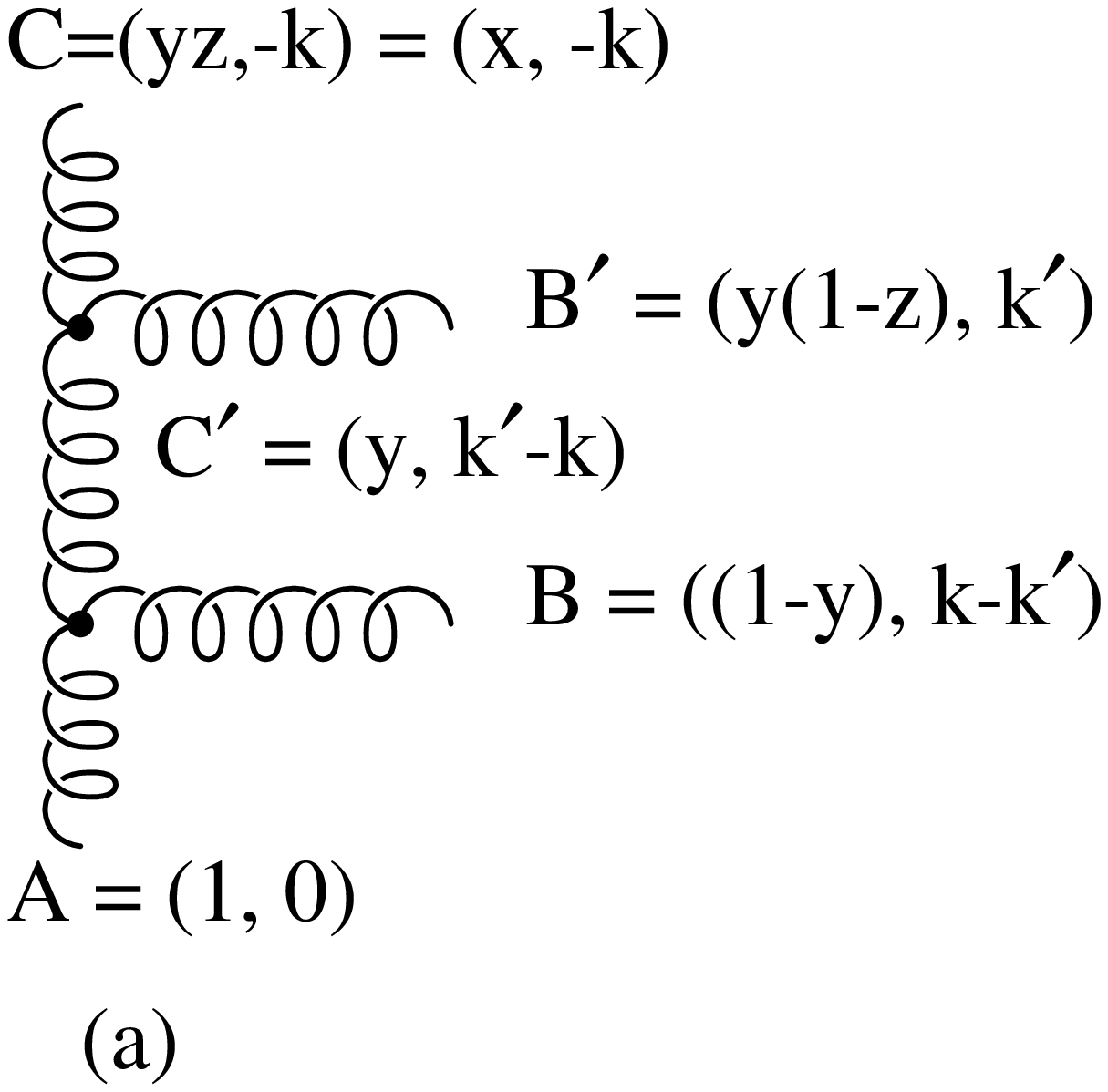}
\hskip -1truecm
\epsfysize=5truecm\epsfbox{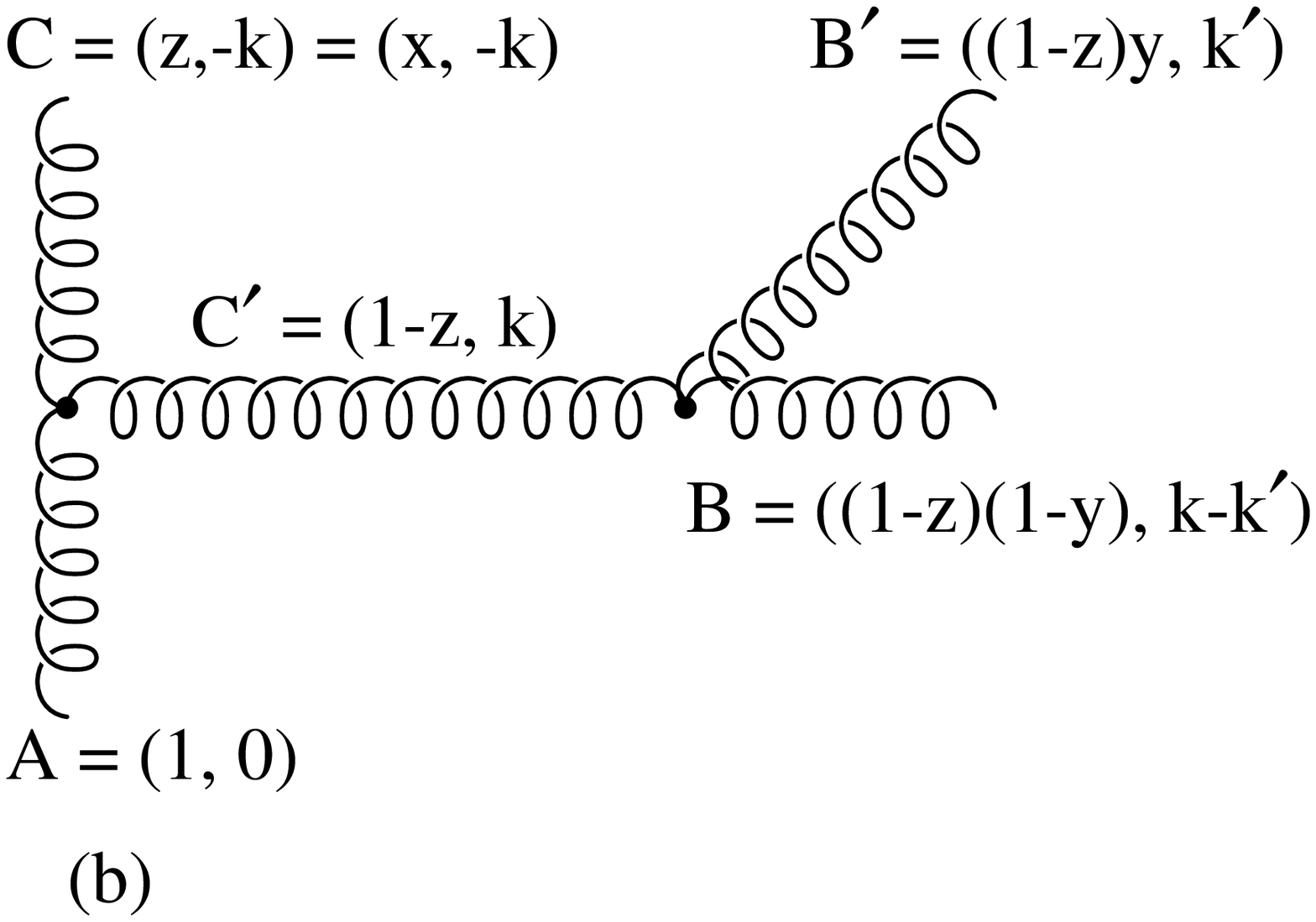}
\hfil}
\bigskip
\bigskip
\noindent{\footnotefont\baselineskip6pt\narrower
Figure 2: Diagrams for (a) t-channel and (b) s-channel gluon emission
contributing to the splitting function. For each line the parton's
energy (in units of $p$) and transverse momentum are indicated.
\medskip}}
\endinsert
\topinsert
\vfill
\eject
\vbox{\vskip 4truecm\hbox{
\hfil\hskip 2truecm \epsfxsize=10truecm\epsfbox{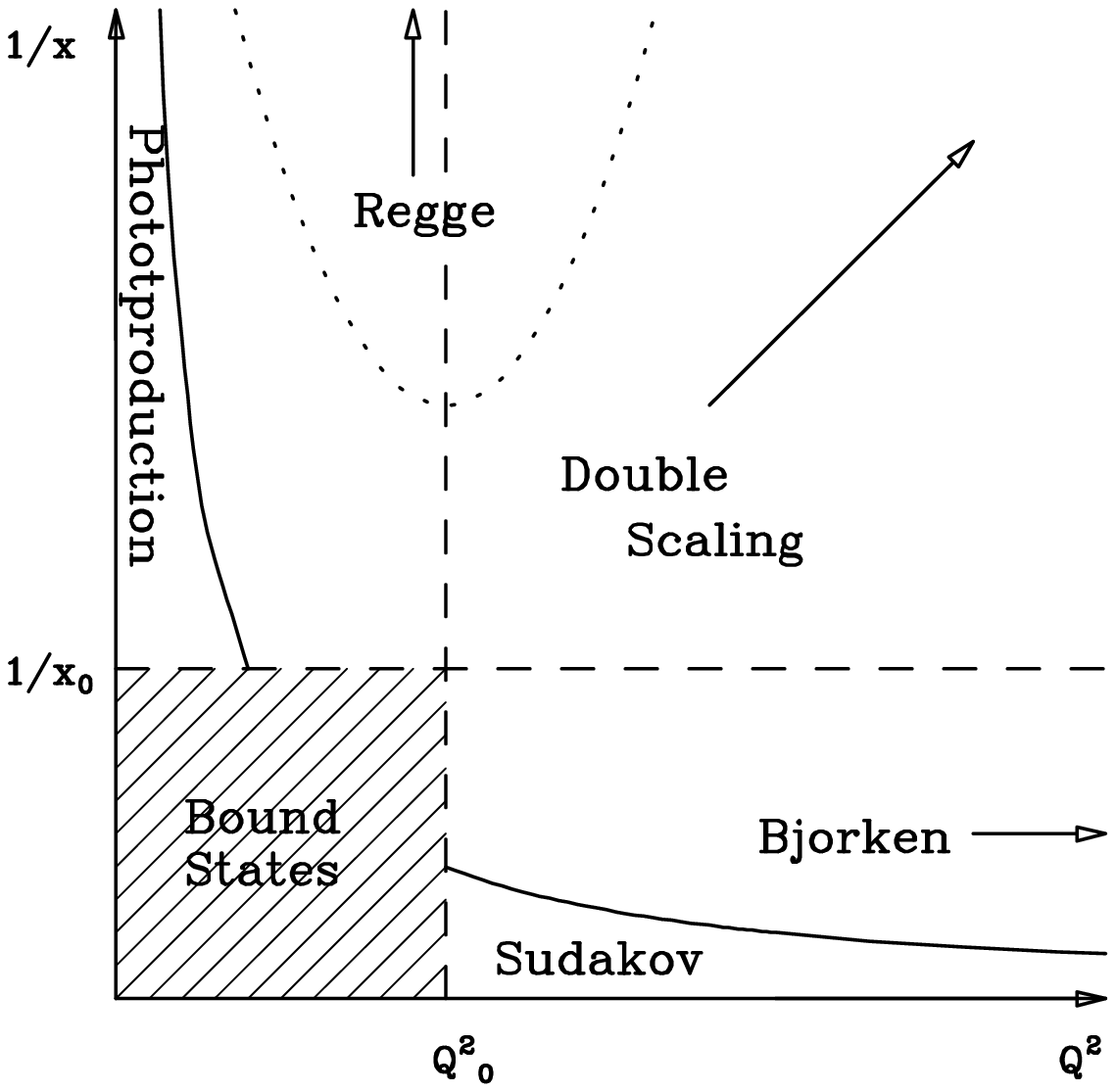}
\hfil}
\bigskip
\bigskip
\noindent{\footnotefont\baselineskip6pt\narrower
Figure 3: Regions and limits in the $(Q^2,{1\over x})$ plane.
The shaded area denotes the region of nonperturbative dynamics where
there is no large scale.
\medskip}}
\endinsert
\vfill
\eject
\bye